\documentclass[fleqn,usenatbib]{rasti}

\usepackage{newtxtext,newtxmath}

\usepackage[T1]{fontenc}

\DeclareRobustCommand{\VAN}[3]{#2}
\let\VANthebibliography\thebibliography
\def\thebibliography{\DeclareRobustCommand{\VAN}[3]{##3}\VANthebibliography}

\usepackage{graphicx}	
\usepackage{amsmath}	
\usepackage{hyperref}
\usepackage[capitalise, noabbrev]{cleveref}  
\crefname{figure}{Fig.}{Figs}
\Crefname{figure}{Fig.}{Figs}
\crefname{table}{Table}{Tables}
\Crefname{table}{Table}{Tables}
\crefname{equation}{equation}{equations}
\Crefname{equation}{Equation}{Equations}
\creflabelformat{equation}{(#2#1#3)}
\crefname{section}{Section}{Sections}
\crefname{appendix}{Appendix}{Appendices}
\usepackage{tikz}
\usetikzlibrary{positioning, fit, backgrounds, arrows.meta}
\usepackage{algorithm}
\usepackage{algpseudocode}
\usepackage{tabularx}

\title[\textsc{OCTAVIUS}: A Galaxy Simulation Analysis Toolkit]{\textsc{OCTAVIUS}: A High-Performance Analysis Toolkit for Galaxy Simulations}

\author[J. P. Duminy, J. Szpila, R. Dav\'{e}.]{
Jean-Paul Duminy,$^{1}$\thanks{E-mail: jp@duminy.org}
Jakub Szpila,$^{3}$
Romeel Dav\'{e}$^{1,2}$
\\
$^{1}$Institute for Astronomy, Royal Observatory, University of Edinburgh, Edinburgh EH9 3HJ, UK\\
$^{2}$Department of Physics and Astronomy, University of the Western Cape, Bellville, Cape Town 7535, South Africa\\
$^{3}$Nicolaus Copernicus Astronomical Center, Polish Academy of Sciences, Bartycka 18, 00-716 Warszawa, Poland.\\
}

\date{Accepted XXX. Received YYY; in original form ZZZ}

\pubyear{\the\year{}}

\begin{document}
\label{firstpage}
\pagerange{\pageref{firstpage}--\pageref{lastpage}}
\maketitle

\begin{abstract}
Increased computing power and more efficient codes have allowed cosmological simulations of galaxy formation to grow massively in scale: the analysis of their outputs has consequently become a big data computational challenge. We present \textsc{Octavius}, a high-performance, hybrid-parallel simulation analysis toolkit. \textsc{Octavius} ingests a simulation snapshot with associated halo catalogue, locates galaxies with a 6D friends-of-friends algorithm, then produces an HDF5 catalogue of properties for haloes, subhaloes, and galaxies. The analysis is fast and memory efficient: we compare the performance against the widely used predecessor toolkit, \textsc{Caesar}, and show \textsc{Octavius} delivers order-of-magnitude improvements in speed with a factor of $>5$ reduction in peak memory. We demonstrate near-linear strong scaling of compute and linear peak memory scaling with the number of halo-assigned particles. Compute times scale with the superlinear computational load of larger haloes; we note the pipeline is presently I/O-limited. The code is written in \textsc{Python} and uses a declarative, plugin-based architecture so as to be easily extensible. \textsc{Octavius} can currently analyse \textsc{SIMBA}, \textsc{SWIFT-KIARA}, \textsc{SWIFT-EAGLE}, \textsc{SWIFT-COLIBRE}, and \textsc{TNG} snapshots; \textsc{AHF}, \textsc{HBT-HERONS}, and \textsc{SUBFIND} halo catalogues are supported. Properties computed include masses, galaxy photometry with dust attenuation, and kinematic, radial, morphological, and virial quantities. Output catalogues are easily accessed through a provided user API. The code is open sourced at \url{https://github.com/jp-duminy/octavius}.
\end{abstract}


\section{Introduction}

In recent years, cosmological simulations have established themselves among the primary avenues for studying the formation and evolution of galaxies \citep{vogelsberger_sim_review}. Simulations serve as the lab of an astrophysicist, where existing theories are challenged and new ones explored, and can reasonably reproduce observational data \citep{vogelsberger_PropertiesGalaxiesReproduced2014,somerville_sim_review}. To reliably produce accurate results, simulations use intricate physical models to model the diverse physical processes which shape galaxies; the length and time scales involved cover orders of magnitude \citep{naabTheoreticalChallengesGalaxy2017}. The growing abundance of available computing power, coupled with improved numerical methods and more efficient codes, have significantly increased the scale of these simulations: modern-day production runs now involve gigaparsec-scale boxes \citep{schaye_FLAMINGO, pakmor_mil_tng}. Such runs take months and produce terabytes of data in the form of snapshots at numerous cosmic epochs along the way. Future simulations are expected to usher in the dawn of exascale computing to astrophysics \citep{shukla_exascale_computing, frontiere_advanced_hydrodynamics}, significantly increasing the output data volume. The increasing scale and complexity of these simulations present a growing big data computational challenge for the analysis and interpretation of their outputs. Scalable, efficient software architecture is necessary to process petabytes of raw simulation data and keep pace with the science demands of current and coming efforts.

Beyond data volume, analysis software comes with many other challenges towards the goal of generating user-friendly catalogues from raw simulation snapshots. There is a diverse range of research interests in the field and, accordingly, a multitude of different simulation suites \citep{dave_SIMBA, schaye_EAGLE, schaye_FLAMINGO, schaye_COLIBRE, vogelsberger_illustris}, based on frameworks such as \textsc{GIZMO} \citep{hopkins_GIZMO}; \textsc{SWIFT} \citep{schaller_SWIFT}; \textsc{AREPO} \citep{springel_AREPO}; and \textsc{GADGET} \citep{springel_GADGET}. Thus, post-processing codes must accommodate many different simulation formats and cater to different user requirements whilst maintaining an agnostic output format.  Furthermore, one of the primary steps in the post-processing of snapshots is the identification of structures such as haloes. This is a well-explored problem which has received a significant amount of attention in the literature \citep{knebe_haloes_MAD, onions_subhaloes_notts, behroozi_mergers_notts, forouharmoreno_HBT}, resulting in the development of many advanced algorithms and halo finder codes over the years. Post-processing codes must therefore support different halo catalogue formats in addition to simulation types.

For over a decade, the analysis toolkit \textsc{Caesar} \citep{thompsonSPHGRSmoothedParticleHydrodynamics2015, dave_SIMBA} has served as the tool-of-choice in post-processing for many simulation suites, including {\sc Mufasa} \citep{daveMUFASAGalaxyFormation2016}, {\sc Simba} \citep{dave_SIMBA}, CAMELS \citep{villaescusa-navarroCAMELSProjectPublic2023}, and The 300 \citep{cuiThreeHundredProject2018}.  \textsc{Caesar} ingests a snapshot at a given cosmic epoch (i.e. redshift), identifies haloes and galaxies, computes a wide range of properties for both, and outputs a reduced galaxy/halo catalogue. However, there are known limitations with its performance. \textsc{Caesar} does not possess the level of efficiency required for fast analysis of modern-day snapshots, and its performance does not scale: its RAM requirements are super-linear to the size of the snapshot, a problem which has become untenable with the advent of the next generation of simulations. Furthermore, the iterative nature of its development has caused the architecture of the code to grow monolithic and entwined, which has complicated maintenance and further development. 

We therefore present \textsc{Octavius}, the appropriately named successor to \textsc{Caesar}. \textsc{Octavius} has been developed from the ground-up with an \textit{ab initio} focus on performance, scalability, and extensibility. Programmatically, this is primarily achieved through \textsc{numba} \citep{Lam_numba}, a just-in-time compiler for \textsc{Python} code which facilitates competitive speed with low-level languages whilst retaining the readability and ease-of-use commonly associated with \textsc{Python}. Moreover, the code is hybrid-parallelised through \textsc{MPI} \citep{dalcin_mpi4py} and the intrinsic multithreading provided by \textsc{numba}. Architecturally, \textsc{Octavius} utilises a modular, declarative `plugin-based' design which is well-suited to extensions and modifications. The code is open-source and permissively licensed under the \textit{BSD 3-Clause Licence}. 

It is not uncommon for scientific software to be confronted with code quality issues which can lead to errors in physical output; this is often attributed to an absence of testing, which is a cornerstone of software engineering \citep{kanewala_testing_software}. The safeguards provided by tests are crucial as the role of LLMs in assisting code development has evolved to support autonomous agentic workflows \citep{obrien_vibecoding}. To remedy this and instil confidence in its analysis catalogues, \textsc{Octavius} features both comprehensive unit testing and a regression suite for the validation of its outputs. The developer environment expands on the unit tests with automatic linting and formatting as part of a continuous integration workflow: a developer manual is provided as part of comprehensive documentation.

The present state of the code contains optimised reimplementations of all \textsc{Caesar} routines. This includes built-in galaxy finding; membership mapping between haloes, galaxies, and particles; a wide assortment of common halo and galaxy properties, including virial, morphological, kinematic, and phase-specific quantities; photometry, which produces apparent and absolute magnitudes with and without dust in all bands supported by FSPS \citep{conroyPropagationUncertaintiesStellar2009, conroy_FSPS}. The analysis produces HDF5 catalogue files which can be interfaced through a provided user-friendly, object-oriented \textsc{Python} API. 

The pipeline is fully configurable through a YAML configuration file. Users can tune physics parameters such as extinction law assumptions or phase thresholds, select which stages of the analysis to include, and make algorithmic decisions. The API also supports post-catalogue analysis reruns such as photometry with rotations and different lines of sight. 

\textsc{Octavius} currently supports \textsc{SIMBA} \citep{dave_SIMBA}; \textsc{KIARA}; \textsc{EAGLE} \citep{schaye_EAGLE}; \textsc{Colibre} \citep{schaye_COLIBRE}; and \textsc{TNG} \citep{vogelsberger_illustris} simulation snapshots.  Other particle-based simulations such as those run with {\sc Gadget} and {\sc Gizmo} should also work out-of-the-box, though they have not been fully tested; this encompasses the vast majority of Lagrangian codes currently being used to run galaxy formation models. Among halo finders, support is provided for \textsc{HBT-HERONS} \citep{forouharmoreno_HBT}, \textsc{AHF} \citep{knollmann_AHF}, \textsc{SUBFIND} \citep{springel_SUBFIND}, and on-the-fly snapshot halo catalogues, again encompassing the most widely used codes at present. The developer's guide provides step-by-step instructions on how to add support for other bespoke codes and halo finders.  The pipeline demonstrates order-of-magnitude improvements in speed and a significantly smaller memory footprint than its predecessor, and efficient scaling with computational load, making it much better suited to handle next-generation galaxy formation simulations.  

The remainder of the paper is organised as follows. In~\cref{sec:description} we first present an overview of the pipeline and a high-level description of the code architecture followed by detailed descriptions of the data model (\cref{sec:data_model}), analysis routines (\cref{sec:randa}), parallelism (\cref{sec:parallelism}), and the user-facing aspects of the package (\cref{sec:api,sec:dandd}).  We validate the core features of \textsc{Octavius} against their counterparts in \textsc{Caesar} in~\cref{sec:validation} and present common scaling relations. Finally, we demonstrate the performance of the code in~\cref{sec:performance}.

\section{Software Description} \label{sec:description}

\textsc{Octavius} was designed not only to remedy the performance issues faced by its predecessor {\sc Caesar} but also to provide a flexible, well-engineered core architecture for future functionality. 

\textsc{Octavius}'s primary feature is its galaxy/halo catalogue generation routine, although it also provides an environment to enable other analyses once the catalogue is generated.  In this section we describe the architecture and implementation of the code, including the algorithms utilised for high performance and parallelism. We further describe the underlying physics behind the more complicated stages of the pipeline.

The 0.9.3 version of the code is described for this section and archived under the \textit{legacy-rasti} branch on the source repository. 

\subsection{Overview} \label{sec:overview}

\begin{figure}
\centering
\begin{tikzpicture}[
    node distance=3.5mm,
    stage/.style={draw, rounded corners, align=center, font=\small,
                  text width=0.58\columnwidth, inner sep=3pt},
    io/.style={draw, align=center, font=\small, text width=0.26\columnwidth},
    flow/.style={-{Stealth}, thick},
    aux/.style={-{Stealth}, dashed},
]
\node[io] (halo_catalogue) {Halo Catalogue};
\node[io, left=6mm of halo_catalogue] (snapshot) {Snapshot HDF5};
\node[io, right=6mm of halo_catalogue] (user_config) {User Config};

\node[stage, below=6mm of halo_catalogue] (assignment)
    {Assign haloes from the catalogue to ranks (\cref{sec:parallelism})};
\node[stage, below=of assignment] (collaborative_read)
    {Ranks use format parsers to collaboratively read data from snapshot and apply halo reduction mask};
\node[stage, below=of collaborative_read] (ingestion)
    {Data stored in efficient, format-agnostic containers (\cref{sec:data_model})};
\node[stage, below=of ingestion] (fof6d)
    {Run the 6D FOF algorithm to locate galaxies in haloes (\cref{sec:fof6d})};
\node[stage, below=of fof6d] (routines)
    {Analysis routines run: load particle data $\rightarrow$ compute physical quantities $\rightarrow$ drop unneeded data (\cref{sec:scheduling})};
\coordinate (lazy_load_corner) at ([xshift=-4mm]routines.west);
\node[stage, below=of routines] (catalogue_write) {Root rank determines the global ordering of the catalogue, then ranks write to output HDF5 file};
\node[io, below=of catalogue_write] (output_catalogue) {Analysis Catalogue};

\draw[flow] (halo_catalogue) -- (assignment);
\draw[flow] (snapshot) |- ([yshift=1.5mm]collaborative_read.west);
\draw[aux] (routines.west) -- (lazy_load_corner)
    |- node[pos=0.25, sloped, below, font=\scriptsize] {lazy load}
    ([yshift=-1.5mm]collaborative_read.west);

\draw[aux]  (user_config) |- (assignment);
\draw[aux]  (user_config) |- (fof6d);
\draw[aux]  (user_config) |- (routines);
\foreach \upper/\lower in {assignment/collaborative_read, collaborative_read/ingestion,
                           ingestion/fof6d, fof6d/routines, routines/catalogue_write,
                           catalogue_write/output_catalogue}
    \draw[flow] (\upper) -- (\lower);

\begin{scope}[on background layer]
    \node[fill=gray!12, rounded corners, inner sep=2mm,
          fit=(collaborative_read)(routines)(lazy_load_corner),,
          label={[font=\footnotesize]west:\rotatebox{90}{per rank}}] {};
\end{scope}
\end{tikzpicture}
\caption{A flowchart of the pipeline. The shaded region represents the section of the pipeline which each rank runs independently (bar the collaborative data reads).}
\label{fig:pipeline_flowchart}
\end{figure}
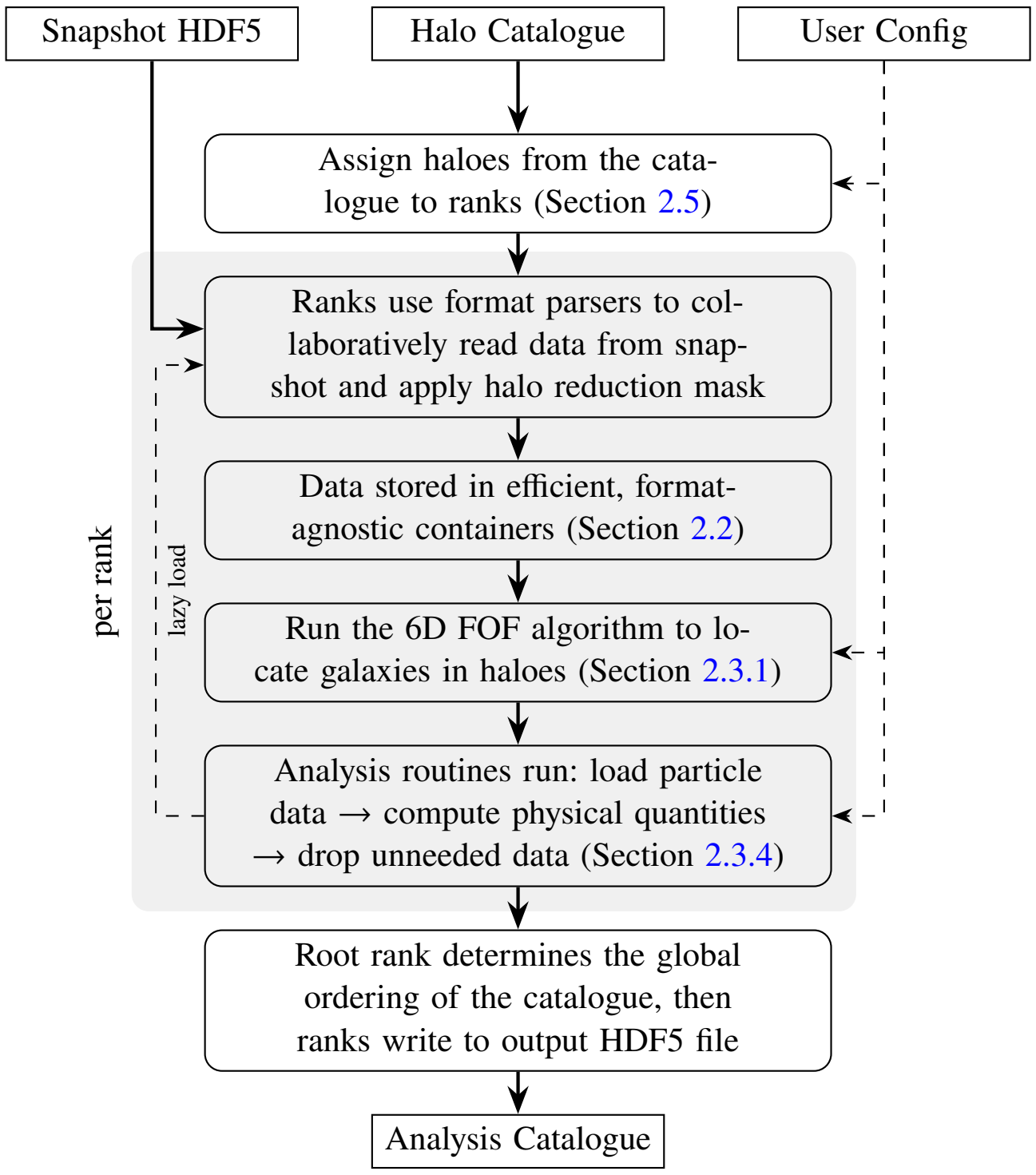

The analysis pipeline is fundamentally a reduction operation: particle datasets are ingested, analysed, and reduced to a catalogue containing the properties of their parent galaxies and haloes. \Cref{fig:pipeline_flowchart} demonstrates the ordering of the pipeline. Under the parallelism model (\cref{sec:parallelism}), each rank is assigned a subset of haloes to process. The 6D FOF algorithm (\cref{sec:fof6d}) locates galaxies within haloes, so the pipeline only needs to consider particles which actually belong to haloes. The first step is therefore to compute a mask for the constituent particles of each rank's haloes, discarding those which either do not belong to a halo or belong to a halo with fewer DM particles than a configurable threshold. This mask is applied to each loaded dataset as a reduction step; commonly, a sizeable fraction of particles ($\approx 40$ per cent at lower redshifts) are not assigned to haloes, so this avoids holding redundant data. 

The codebase is architecturally declarative. This means the analysis routines are fully decoupled from data transport: they simply declare what they need from the snapshot. The loading mechanism will then lazily load the required datasets into the internal data structures which the routines operate on. Each routine computes a subset of related properties derived from similar datasets. For example, the core property routine derives kinematic, virial, and morphological quantities from all particle positions, velocities, potentials, and masses; the photometry routine uses gas and star-specific datasets such as metallicities and star formation rates to compute mock photometry. Since properties only depend on the constituent particles, haloes (and consequently galaxies) are processed independently. The pipeline is thus embarrassingly parallel, meaning the outputs of each routine are always invariant to the number of ranks and/or threads. 

Physical and algorithmic parameters, alongside pipeline-related options such as which particle types to analyse and which routines to run, are configured by the user through a single YAML file. The introduction of a configuration file provides two advantages over \textsc{Caesar}: users have more control over the analysis, and the configuration and declarative architecture are used in tandem to validate the pipeline at instantiation (\cref{sec:scheduling}).

On each rank, the routines write their outputs into type-specific galaxy/halo data structures. The tendency is that users will want to reanalyse the same snapshot multiple times, hence the analysis catalogue should be deterministic. Once the analysis has finished, the root rank gathers group-level information which is used to determine a global ordering of the data. The ordering is by mass, with tiebreaks settled by unique identifiers: this achieves determinism on the same configuration. Once the order is broadcast to the other ranks, the root rank creates an output HDF5 file to which all ranks write their data. 

The output file will contain a catalogue of halo and galaxy properties. Crucially, this includes membership relationships such as which galaxies are centrals or satellites; the subhalo hierarchy; mappings back to the original halo/merger tree catalogue; and particle membership arrays. Thus, \textsc{Octavius} catalogues enable users to explore the relationships between structure in the snapshot and reaccess constituent particles to compute any desired properties beyond the many already produced by the pipeline. The separation of concerns between analysis and the data model means the catalogue is fully agnostic to the input format, so \textsc{Octavius} can be used to directly compare the outputs of different simulation suites. This is made easier by the provided object-oriented API (\cref{sec:api}) for loading and working with catalogues.

\subsection{Data Model} \label{sec:data_model}

The transport and representation of data at scale is among the primary challenges faced by scientific software. Research groups will naturally focus their programming efforts towards examining the physics: as such, the data model is often treated as a means to an end and is tightly coupled to the simulation format. Such software is complicated to extend, refactor, or reuse with different simulations. Furthermore, the scales of modern simulations effectively enforce parallel programming as a standard, rather than an optimisation; few researchers are trained to parallelise code.

The data model of \textsc{Octavius} is designed to directly address these concerns. Through separation of responsibilities, the analysis routines can operate on format-agnostic data. This is achieved through an ingestion layer of format-specific parsers for each supported simulation suite. An example of where this is useful is with regard to hydrogen mass fractions: simulations like \textsc{SIMBA} implement on-the-fly self-shielding following the \citet{rahmatiEvolutionColumnDensity2013} prescription, whilst other simulations such as \textsc{TNG} require a bespoke treatment for star-forming gas (cf.~\citet{stevensAtomicHydrogenIllustrisTNG2019}, Appendix A1). Their respective parsers separately account for this and produce a consistent definition of the hydrogen mass fractions. The codebase also maintains internal code units and data types, the conversions of which are backed by \textsc{astropy} \citep{astropy_3, astropy_2, astropy_1} to prevent unit mismatch errors. 

The parsers inherit a base interface which handles interfacing with HDF5 files and \texttt{MPI} logic. Extending the code to support a new simulation format therefore requires no familiarity with parallel programming and simply involves massaging the format conventions into the expected agnostic form.

An abstraction layer exists between the simulation parsers and analysis routines in the form of efficient, purpose-built data structures for handling in-memory data. The \textsc{Octavius} pipeline is a reduction operation: particle-level snapshot data is aggregated into a catalogue of halo and galaxy properties. Both particle and group-level data are therefore separately stored in lightweight, type-specific containers. The containers for galaxy and halo data store mappings to their constituent particles in compressed sparse-row (CSR) format. This enables quick $\mathcal{O}(1)$ lookups into the corresponding particle containers for efficient property aggregations. Since computing the properties of a group only requires accessing its particles, the analysis routines process each group independently. This forms the basis of the embarrassingly parallel strategy described in~\cref{sec:parallelism}. The lookup-based storage approach also enables a novel, useful feature not present in \textsc{Caesar}: inclusive properties, such as those of a parent halo from its subhaloes, can be computed by propagating the index pointers to the subhalo particles into the membership mapping of its parent halo.

As specified in~\cref{sec:overview}, the code is architecturally declarative such that each analysis routine is structured as a self-contained stage of the pipeline. Routines declare their required datasets, dependent stages, and outputs in an internal YAML file. Self-describing routines can be validated at instantiation to catch inconsistencies and bugs early. The data model uses the YAML file to determine when to load and release data, thereby decoupling the routines from the parsers. This significantly simplifies extending the code for new properties: adding an analysis routine involves modifying a YAML file and writing physics logic rather than handling data transport. 

\subsection{Routines and Algorithms} \label{sec:randa}

In this section we outline and describe the \textsc{Octavius} analysis routines along with the algorithms utilised to improve scalability and performance. This includes the scheduling of the pipeline.

\subsubsection{Friends-of-friends} \label{sec:fof6d}

The identification of haloes themselves in simulations is a non-trivial task for which a variety of well-established finder codes exist \citep{knebe_haloes_MAD, onions_subhaloes_notts, behroozi_mergers_notts, forouharmoreno_HBT}. These finders typically locate bound DM overdensities to which they then assign baryons. Separating galaxies from their DM hosts is not as straightforward as only considering the baryonic component: galaxies are surrounded by a diffuse stellar halo possessing different kinematic characteristics. There is also a distinction between gas in the interstellar medium (ISM) and circumgalactic medium (CGM), where gas in the former is cold, dense (and often star-forming), and gas in the latter is warm and diffuse. \textsc{Octavius} therefore runs the friends-of-friends (hereinafter FOF) algorithm \citep{davis_FOF} in haloes to find galaxies by locating baryonic phase-space overdensities with ISM-like gas.

The 3D FOF algorithm defines a characteristic linking length $\ell_{\rm FOF} = b \bar{\lambda}$. $\bar{\lambda}$ represents the mean comoving baryonic interparticle separation. $b$ is a configurable dimensionless scaling parameter, usually taken to be 0.2 when locating haloes; galaxies are comparatively more dense, motivating a lower default value of 0.02. For reference, $b = 0.02$ corresponds to a linking length of $\approx$ 2.9 $\mathrm{c \, kpc}$ in the fiducial \textsc{SIMBA} snapshot. To distinguish the ISM and CGM components, gas particles entering the algorithm are subject to a configurable criterion on their temperature, density, and star formation rate. Baryonic particles are then linked to neighbours within $\ell_{\rm FOF}$ to produce FOF groups. A well-known limitation of 3D FOF is its potential to artificially link distinct structures when they are bridged together by filaments or overlapping (such as in merger events). Consequently, following \citet{diemand_FOF}, we extend the algorithm to phase space (6D FOF). 6D FOF is also useful to trim `flyby' particles which spatially coincide with galaxies. Rather than normalising to the velocity dispersion of the FOF group, we adopt the localised velocity criterion used by \textsc{Caesar}: for each particle $i$, we compute a kernel-weighted RMS relative velocity $\sigma_i$ of its configuration-space neighbours as in~\cref{eq:fof6d_vel}.

\begin{equation}
    \label{eq:fof6d_vel}
    \sigma_i^2 = \frac{\sum_{j \neq i} W_{ij} \lvert \mathbf{v}_{ij} \rvert^2}{\sum_{j \neq i} W_{ij}}
\end{equation}
where $W_{ij}$ is the cubic spline kernel described in \citet{monaghan_sph} with support radius $\ell_{\rm FOF}$, and $\mathbf{v}_{ij} = \mathbf{v}_{i} - \mathbf{v}_{j}$. 

As $\sigma_i$ depends only on the neighbours within $\ell_{\rm FOF}$, the velocity criterion applied this way represents a simple extension to the spatial search which requires no prior knowledge of group membership. Particles are then linked in velocity space if $\lvert \mathbf{v}_{ij} \rvert \leq \alpha \max ( \sigma_i , \sigma_j)$. $\alpha$ is another configurable dimensionless scaling parameter, taken to be 1.0 to match its value in \textsc{Caesar}. The choice of the larger dispersion (i.e. the more permissive environment) is intended to avoid over-fragmentation of structure. 

Connected components in phase space are identified through a path-compressed union find. These components are only considered galaxies if they pass a configurable minimum star particle threshold; this is based on number rather than mass so as to be tunable to the numerical resolution of the simulation, and is set to 16 by default. The default parameters of the aforementioned gas filter correspond to gas with a hydrogen number density of $n_{\rm H} > 0.13 \, \mathrm{cm^{-3}}$ that is either cool ($T < 1 \times 10^5 \, \mathrm{K}$) or star-forming. In~\cref{sec:validation} we demonstrate the 6D FOF produces a sensible galaxy population with expected scaling relations.

In the implementation of the algorithm, we exploit the fixed value of $\ell_{\rm FOF}$ to construct a spatial hashing grid. This approach is similar to that described in \citet{creasey_FOF}: particles are organised into cells sized by $\ell_{\rm FOF}$ such that identifying the neighbours of a particle only queries 27 cells. \citet{creasey_FOF} reports spatial hashing can produce speed-ups of up to a factor of 20 over more commonly used tree structures such as those described in \citet{feng_FOF}. Particles in haloes are centrally concentrated: since the grid is sized to the extent of a halo, most cells are empty. The grid is therefore stored in a sparse format (only storing occupied cells), thereby avoiding redundant memory allocation. Periodic boundary conditions are handled through the minimum-image convention for each halo, which is valid provided the extent of a halo does not exceed half of the box size.

The construction of the grid requires sorting particles by cell index, which has time-complexity $\mathcal{O}(N \log N)$. The pair evaluations scale as $\mathcal{O}(N \bar{k})$ where $\bar{k}$ is the average number of candidate neighbours: as $\ell_{\rm FOF}$ is fixed, $\bar{k}$ is density-dependent, and can tend to $N$ in sufficiently dense regions. The union finds incur an effectively constant cost when averaged over all operations \citep{tarjan_worstcase}. Pairs are evaluated then discarded, resulting in a memory-complexity of $\mathcal{O}(N)$. 

\subsubsection{Photometry}

The James Webb Space Telescope (JWST) has advanced our understanding of galaxy formation and evolution through its detailed, novel observations at high redshift \citep{gardnerJamesWebbSpace2023}. At the same time, modern simulations have reached a level of sophistication where they can now confront observational data \citep{vogelsberger_sim_review}. This has caused a surge in demand for mock photometry \citep{wilkins_FirstLightReionization2023}. However, creating SEDs of even a single galaxy using full radiative transfer is computationally expensive \citep{narayananPOWDERDAYDustRadiative2021}. \textsc{Octavius} includes a fast, particle-based photometry routine which accounts for dust and can quickly, reliably provide population statistics such as luminosity functions; this is an optimised rework of its equivalent in \textsc{Caesar}.

This is done by first computing the dust extinction to each star particle from the kernel-weighted metal column density along the line of sight. The kernel integral takes the form in~\cref{eq:kernel_los}

\begin{equation}
\label{eq:kernel_los}
    \Sigma(b) = 2 \int_{0}^{+\infty} W\!\left(\frac{\sqrt{b^2 + z^2}}{h}\right) \, dz
\end{equation}

where $b$ is the impact parameter, $z$ the coordinate along the line of sight, and $W$ the kernel weight with support radius $h$ where $h$ is the smoothing length. This integral is precomputed across tabulated values of $b \, / \,h$ via trapezoid integration for speed; the granularity of the tabulated values is configurable. We then proceed to derive the metal column density via ray tracing to each star. 

When ray tracing to each star we only consider the contribution of gas in its parent galaxy: this approximation is on the basis that the CGM gas is diffuse and makes a minor contribution to the extinction relative to the considerable computational expense its inclusion would incur. This does also pose a limitation as it excludes other galaxies along the line of sight. To sum the metal column density we must consider the gas particles which our ray intersects. A na\"{i}ve approach would be to check this condition for each gas particle. This is $\mathcal{O}(N_{\rm gas} \times  N_{\rm star})$ in time. We therefore adopt an almost identical approach to the 6D FOF described in~\cref{sec:fof6d}. For each galaxy, we construct a cell grid for its gas; the cells are sized to the maximum smoothing length $h_{\rm max}$ of any gas particle in the galaxy. Beyond $h_{\rm max}$ the kernel support is zero which sets $W = 0$ in~\cref{eq:kernel_los}; therefore, provided $h_{\rm max}$ is not of similar size to the galaxy itself, the cell partitioning will reduce redundant checks against gas particles for every star. The grid described in~\cref{sec:fof6d} was stored in a sparse representation to account for the central concentration of haloes: however, galaxies are found to be dense enough that almost all cells are usually occupied, rendering the sparse representation unnecessary. Following the construction of the grid, the metal column density along the line of sight is computed from~\cref{eq:sigma_z}

\begin{equation}
\label{eq:sigma_z}
    \Sigma_Z = \sum_{j} \frac{m_j Z_j}{h_j^3} \, \Sigma(b_j)
\end{equation}

where the index $j$ represents gas particles in front of the star. $m_jZ_j$ represents the metal masses of the gas particles: for simulations with a dust model (e.g. \textsc{SIMBA}), these are replaced by the stored dust masses. 

This leads to the value of $A_{V}$ for each star from \citet{watsonGalacticDusttometalsRatio2011} as in~\cref{eq:av}

\begin{equation}
\label{eq:av}
    A_V = \frac{X_{\rm H} \, (1+z)^2 \, \Sigma_Z}{m_{\rm p} \, (N_{\rm H}/A_V) \, Z_{\sun}}
\end{equation}

where $\Sigma_Z$ is the metal column density (comoving, converted to physical units by the factor of $(1+z)^2$); $X_{\rm H}$ is the primordial hydrogen fraction (0.76); $m_{\rm p}$ the proton mass; and $N_{\rm H}/A_V = 2.2 \times 10^{21}\,{\rm cm^{-2}\,mag^{-1}}$ and $Z_{\sun} = 0.0189$ are adopted from \citet{watsonGalacticDusttometalsRatio2011}. For simulations with no dust model, a metallicity-dependent dust correction from \citet{liDusttogasDusttometalRatio2019} is applied.

The spectra of each star (with and without dust attenuation) are then generated by interpolating its metallicity and age onto a precomputed SSP table (configured by the user) generated by \textsc{FSPS} \citep{conroy_FSPS}. Two options are included to improve the accuracy of the spectra. Firstly, star particles represent a population of stars but carry a single formation time: this matters for young stars whose SEDs vary considerably in their infancy. To remedy the coarse sampling of young stars by the resolution of the simulation, we split the ages of particles below a configurable age threshold into evenly spaced sub-populations. The sub-populations are then interpolated individually. Second, the option to oversample the SSP table is included to improve the accuracy of the interpolated spectra. 

The dust-corrected spectrum of each star is then attenuated by $A_\lambda = A_V \, k(\lambda) / k(V)$ according to a configurable choice of extinction law. The Doppler shift from the peculiar velocity is then applied to both the dust-corrected and uncorrected spectra; both spectra are then added to their counterparts in the parent galaxy. The galaxy spectrum can then itself be optionally attenuated for cosmic extinction as prescribed in \citet{madauRadiativeTransferClumpy1995}. Users can specify which filters they want photometric properties in; all filters supported by \textsc{FSPS} are available. Properties are then computed in these filters by applying the corresponding bandpasses.

The novelty of the implementation of the routine in \textsc{Octavius} is the cell grid structure for improved performance and scaling. The ground-up reimplementation of the routine allowed the code to be restructured so as to minimise memory allocations and take advantage of the improved data model (\cref{sec:data_model}), which has afforded further performance improvements. This is reflected in the significant speed improvement of \textsc{Octavius} over \textsc{Caesar} demonstrated in~\cref{sec:performance}. The previous runtime dependency on \textsc{FSPS} was avoided by additionally moving the filter transmission data into the HDF5 data file which stores the SSP table: this avoids any \textsc{Fortran} bindings in the pipeline and requires negligible extra storage.

\subsubsection{Other Routines}

In addition to galaxy finding and photometry, \textsc{Octavius} contains three other analysis routines which between them compute a wide assortment of common galaxy and halo properties. These were formerly one unified group property routine in \textsc{Caesar}. However, under the architecture of the new code (\cref{sec:overview}, and further detailed in~\cref{sec:scheduling}), they have been modularised according to the snapshot datasets each routine requires. All \textsc{Caesar} properties have been carried forward, with minor additions.

The first of the three is the core property routine, which is divided into smaller subroutines and computes many fundamental group properties derived from their particles' basic phase-space and mass information. Each subroutine computes a subset of properties which are all derived from the same particle datasets, folding several properties into one function so as to avoid redundant memory allocations or recomputations. Kinematic, radial, virial (for haloes), and morphological (for galaxies) properties are modularised this way. It is sometimes of interest to consider properties at the level of particle types (e.g. the separate velocity dispersions of stars and gas), thus each property (bar virial quantities) is computed for both individual particle types and overall aggregations (baryonic and total). 

The second is the particle-type-specific routine, which contains a subroutine for each baryonic particle type. This handles type-specific properties such as Eddington fractions for black holes, SFR-weighted metallicities for gas, or mass-weighted ages for stars. These reductions require specific particle-level datasets which are not always present in each snapshot (e.g. in DM-only simulations), hence the modularisation of this routine is necessary for format agnosticism.

The third is the local environment routine, which computes aperture masses in configurable aperture sizes for galaxies; this is taken by default to be $30 \, \rm{ckpc}$. Aperture masses additionally consider particles not assigned to galaxies (including DM); $\rm HI$ and $\rm H_2$ apertures are also available. They are computed through the construction of a k-d tree (using \textsc{scipy} \citep{virtanenSciPy10Fundamental2020}) on the parent halo, with spatial queries made from the centre-of-mass of each galaxy. This imposes a limitation, carried over from \textsc{Caesar}, in that particles unassigned to the halo but in the local environment will not be included in the aperture.

\subsubsection{Pipeline Scheduling} \label{sec:scheduling}

The running order of the analysis routines, and the loading and release of their requisite datasets, are automatically determined using the declarative architecture described in~\cref{sec:data_model}. This ensures self-consistency, eliminates redundant I/O, and simplifies extensions to the code.

Routines declare both the datasets they require from the snapshot and any other routines which they require outputs from. In practice this cross-routine dependency is minimal, e.g. aperture masses from the local environment routine require centre positions, which come from the core kinematic routine. Independently, each output column which can appear in the catalogue is declared with associated metadata which includes units and comoving scale factor exponents.

At runtime, the YAML file where this is declared is first validated. This includes verifying all output columns are produced by one routine and every routine's outputs are declared. The self-consistency of the pipeline is thus validated at instantiation, preventing associated bugs from surfacing in output catalogues or during routines.

Users can freely toggle routines. Therefore, the first step of the pipeline scheduling is to ensure all dependencies of the requested routines are themselves enabled. This involves recursively checking over the enabled routines and enabling their dependencies. The running order must then be sorted such that routines run after their dependencies. The declarations mean this can be achieved by representing the pipeline as a simple directed acyclic graph, which is topologically sorted with Kahn's algorithm \citep{kahnTopologicalSortingLarge1962}. The topological sort provides another consistency check: it catches any dependency cycles in the pipeline at instantiation. \texttt{MPI} ranks execute routines in the same order to eliminate desynchronisation issues.

Users can also specify which particle types are included in the analysis. In the case of a dark-matter simulation, a stage like photometry should not run. Therefore another consistency check is performed where the requested routines are checked against which particle types are present in the snapshot and included by the user: if a routine is enabled but none of the particle types it requires are present, it and its dependents are disabled and a warning is emitted.

The pipeline then executes the routines in the sorted running order. Before each routine, the ingestion layer checks which requisite datasets already exist in the internal data structures, then loads the rest. After the routine has finished, the pipeline checks which present datasets are no longer required and drops them. Thus, each dataset is loaded once at most; this reduces the I/O cost of the pipeline. Dropping no-longer-required datasets reduces the resident memory footprint to only the datasets required by the current and remaining routines. 

The declarative architecture allows the pipeline to validate its self-consistency and improve its efficiency. Most importantly, it significantly simplifies extensions to the code: when adding new routines, simply declaring their inputs and outputs means the requisite data will be loaded and outputs written to the HDF5 catalogue automatically. 

\subsection{Optimisation}

\textsc{Python} has become the \textit{lingua franca} of scientific computing owing to its ease of use and readability. The language has developed a strong idiom of vectorisation through \textsc{numpy} \citep{harrisArrayProgrammingNumPy2020}. However, compound \textsc{numpy} operations make intermediate memory allocations which can be costly at scale and unfriendly to memory-intensive work. Furthermore, the per-group reduction operations performed by the \textsc{Octavius} analysis routines use variable-length segments of the particle data (\cref{sec:data_model}). This is supported in basic form by \textsc{numpy} through the \texttt{reduceat} method on its universal functions; however, many of the specific operations in the analysis routines do not straightforwardly vectorise. For example, attenuating star particles in photometry requires integrating the metal column density from gas along the line of sight for each star: vectorisation would require allocating intermediates for each star against all gas which would incur a quadratic cost in memory. Peak memory usage determines whether a snapshot can be analysed for a given hardware configuration; it is therefore imperative to avoid unnecessary allocations.

\textsc{numba} is a just-in-time (JIT) compiler for \textsc{Python} code which offers comparable speed to low-level languages for numeric operations and supports compilation for much of the \textsc{numpy} library \citep{Lam_numba}. Computations written in \textsc{numba} can use explicit loops to perform the necessary reductions whilst avoiding intermediate allocations. Moreover, the package provides built-in multithreading. All of the core physics computations in \textsc{Octavius} are therefore written in \textsc{numba}. Traditionally, performance-intensive codes route this work to compiled \textsc{Fortran} or \textsc{C} extensions. However, this \textsc{Python}-based approach offers a similar level of speed and lowers the barrier to entry for developers unfamiliar with low-level languages who want to add performance-intensive routines to \textsc{Octavius}; without compiled extensions, the package has no build step. The caveat to using \textsc{numba} is the JIT compilation cost; however, this takes all of a minute when first running an analysis, and thereafter the compiled binaries are cached.

The data structures are purpose-built for reduction operations. Each particle quantity is stored as a contiguous array, rather than storing an array of properties for each particle. This means functions only access the particle data they need. The stored CSR membership arrays mean functions can thread over groups (via the offsets array, \cref{sec:data_model}) and access the properties of their particles without making copies. Furthermore, the compute functions are themselves designed to minimise allocations. A straightforward example is in kinematics. Angular momentum, kinetic energy, the inertia tensor, and the mass-weighted velocity dispersion all rely on per-particle positions and velocities relative to the group centre: all four properties are therefore computed in one function, as opposed to individual functions for each with either an allocation for relative positions and velocities or recomputation thereof.

\subsection{Parallelism} \label{sec:parallelism}

Parallel programming is invaluable when it comes to performance-intensive codes, but often imposes a costly engineering challenge. Spatial decomposition is a common approach and usually involves exploiting format-specific conventions such as Peano-Hilbert ordering of data on disc. However, the goal of parallelism is to evenly divide the work amongst processors; this aim is often in conflict with the large density variations within snapshots, necessitating more programming effort and abstraction to balance the workload. 

Task-based parallelism is consequently a more appealing approach, provided there is a simple method of weighting the tasks such that they are equal. \textsc{Octavius} is well-suited to this strategy: its analysis routines process haloes independently and are halo-local. Therefore, each halo can be viewed as a self-contained unit of work. Programmatically, this renders the pipeline embarrassingly parallel, which eliminates the need for inter-rank communication during the routines and is considerably easier to engineer. No assumptions are required regarding the data layout. 

However, haloes are not identical units of work. They contain between tens and millions of particles of different types, which prevents a na\"{i}ve approach where 
each rank claims an equal fraction of the total number of haloes. We therefore seek a means of quantifying the computational weight of a halo such that the distributed tasks produce a balanced runtime. This is possible by heuristically considering the underlying physics: we expect the load to scale superlinearly and for the high end of the halo mass function to dominate the runtime. Therefore, each stage defines a physically and algorithmically motivated power law to describe the weight of a halo as a function of its composition. 

\begin{figure}
    \centering
    \includegraphics[width=\columnwidth]{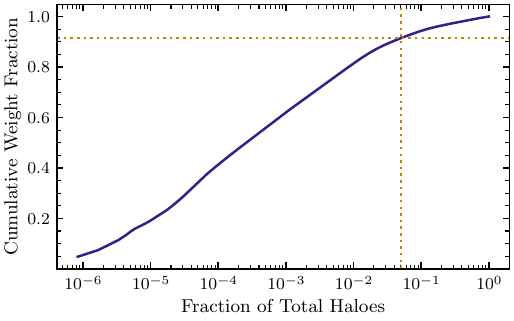}
    \caption{The cumulative weight profile of the haloes on the fiducial \textsc{SIMBA} snapshot according to the power laws (all routines). The yellow dotted line marks the contribution from the top 5 per cent of haloes, which are shown to comprise 92 per cent of the total modelled computational weight. Notably, the single most massive halo can be seen to occupy $\sim$ 5 per cent of the total weight by itself.}
    \label{fig:halo_weight_profile}
\end{figure}

The three basic property analysis routines have no preference for a particular type of particle and limited algorithmic scaling, and thus scale simply as $N_{\rm tot}$. By contrast, the galaxy finder only considers baryonic particles and applies a filter to gas particles to exclude its warm, diffuse CGM component: we would expect this routine to primarily scale with stars with a lesser contribution from gas. Black hole particles represent a negligible fraction of total particles and therefore need not be considered. The galaxy finder power law is then defined as $N_{\rm star}^{1.2} + N_{\rm gas}$. The photometry routine is computationally intensive and only considers stars and gas. Its law is thus defined as $(N_{\rm star} + N_{\rm gas})^{1.1}$. These power laws received minor tuning during development but did not stray far from their initial guesses: the exponents were primarily adjusted downwards to avoid overshooting on single large haloes. 

The resulting weight distribution is shown in~\cref{fig:halo_weight_profile}, where it is evident a minority (5 per cent) of haloes comprise the majority (92 per cent) of the weight. An interesting feature of this plot is at the low end: a single halo, the most massive, accounts for $\sim$ 5 per cent of the weight alone. On zoom simulations, where a specified region of the box has a much higher resolution, a halo could exist which is disproportionately large relative to the others. We note this as a potential, as-of-yet unencountered limitation.

To distribute haloes such that the runtime across ranks is balanced, we employ a straightforward longest-processing-time-first algorithm. The weight of each halo above the user-specified minimum dark matter particle threshold is calculated by summing its weight in each enabled routine: haloes are then assigned to ranks according to~\cref{alg:binning}.

\begin{algorithm}
\caption{The algorithm used to distribute haloes such that the runtime of the pipeline is balanced across ranks.}
\label{alg:binning}
\begin{algorithmic} 
    \State \textbf{Set} all rank\_weights $= 0$
    \State \textbf{Set} all rank\_haloes $= \emptyset$
    \State \textbf{Order} all halo\_weights descending
    \For{each halo $h$}
        \State $i \leftarrow \texttt{argmin(rank\_weights)}$
        \State $\texttt{rank\_weights}[i] \mathrel{+}= \texttt{halo\_weights}[h]$
        \State $\texttt{rank\_haloes}[i] \leftarrow \texttt{rank\_haloes}[i] \cup \{h\}$
    \EndFor
\end{algorithmic}
\end{algorithm}

In~\cref{sec:performance} we demonstrate this simple, physically motivated domain decomposition approach accurately captures the underlying computational weight distribution to produce well-balanced rank runtimes.

The root rank orchestrates the binning algorithm and broadcasts the global allocation to the other ranks. Datasets are then read from the snapshot collaboratively: each rank reads a contiguous chunk of the dataset then participates in a collective \texttt{MPI\_Alltoallv} call to redistribute particles to the other ranks according to the global allocation. The ranks also mask their local particles to include only those belonging to their assigned haloes; as the binning algorithm does not consider haloes below the user-specified minimum dark matter particle threshold, the snapshot reduction step is applied here naturally.

Intra-rank multithreading follows directly from the embarrassingly parallel nature of the pipeline. Threads independently process haloes in parallel by accessing their particles through the aforementioned CSR membership arrays. The intra-rank parallelism is provided through \textsc{numba}.

\subsection{User API} \label{sec:api}

The user API of \textsc{Octavius} is designed in light of the fact \textsc{Caesar} served users with a diverse range of programming experience, ranging from students with basic \textsc{Python} familiarity to researchers proficient in programming. The API is therefore intended to be simple and easy to learn whilst offering flexibility and depth for users seeking full control over the code. 

The parameters and physical assumptions used in each analysis routine (e.g. the temperature above which gas is considered hot or the choice of extinction law) are configurable through a YAML file. This configuration file also grants control over which routines to run and which particle types to analyse. The pipeline is natively MPI-aware and thus the only parallelism-related configuration choice is how many cores to allocate to each process. 

There are two entry points to the pipeline: a command line interface and a Python function. By providing both entry points, \textsc{Octavius} can either be embedded in Python scripts (which is convenient for more interactive workflows) or called from dedicated analysis jobs on clusters. The YAML file is intended to be the primary input to both. To avoid generating many nearly identical files or repeatedly modifying an existing one, both entry points allow parameters such as the snapshot and halo catalogue file paths to be overridden; this supports workflows which involve iterating over snapshots. In Python, the \texttt{OctaviusConfig} class is exposed to the user, allowing the parameters to be passed individually as arguments. As such, the YAML file can be entirely circumvented in Python if desired. The parameters used to generate each catalogue are stored as attributes of the HDF5 file and include any overrides. 

The \textsc{Octavius} API provides rich support for working with the output HDF5 files and enables users to further compute any desired property for groups. The API allows each file to be loaded into an object-oriented catalogue interface. This is known as an \texttt{OctaviusCatalogue} class which contains methods such as \texttt{get\_dataset} that handle the HDF5 file internally. This way, datasets can be loaded lazily with unit conversions or masks. Furthermore, users can access the full membership hierarchy: haloes are mapped to their child subhaloes and galaxies, and all groups store global index mappings of their constituent particles into the original snapshot. The group-particle index mappings enable users to compute new group properties from the snapshot datasets. All of this information can be accessed easily through the \texttt{OctaviusCatalogue} object; however, for flexibility, the HDF5 file itself is designed to be easy to access directly with \textsc{h5py}. 

Though \textsc{Octavius} is primarily intended as a catalogue builder, the API includes the \texttt{OctaviusAnalyser} object for `post-catalogue' analysis. The \texttt{OctaviusAnalyser} uses an \texttt{OctaviusCatalogue} and its associated snapshot to rerun analysis routines on a small subset of the catalogue. Using the global particle index mappings in the catalogue, only the particles belonging to the selected groups are read from the snapshot. Post-catalogue analysis is intended for users who wish to explore the properties of individual groups in the catalogue: the primary use case is photometry, where the effect of orientation on the magnitudes of a galaxy can be investigated. This is a new feature designed to support planned future routines such as line absorption spectra, datacubes, and light cones which are currently in development.

\subsection{Deployments \& Dependencies} \label{sec:dandd}

Unlike its predecessor, \textsc{Octavius} features a designated production environment with deployments to package managing services \texttt{PyPI} and \texttt{conda-forge}. The deployments make it easier for users to install the package; the distinction between production and developer versions prevents users from dealing with untested, experimental features.

Deployments are triggered automatically when new releases are published. To be published, a new release must pass a standard continuous integration workflow: this runs all unit tests, lints the codebase, and checks for issues with the documentation such as link rot. Contributors are also expected to separately run the regression tests on their own larger snapshots to demonstrate any changes in performance or the catalogue outputs. Once the release passes these criteria, it will be available to users within hours. The continuous integration workflow is designed to act as a safety net to prevent bugs or physical errors from reaching users. 

Testing and linting are also applied to pull requests, which additionally receive a testing coverage report. This is to ensure the testing surface of the codebase does not degrade over time. Testing results and the coverage of the current codebase release are transparently published: at present, the coverage is 92 per cent. 

The documentation is hosted on \texttt{readthedocs} and includes both user and developer manuals. The source code itself is also thoroughly documented.

The codebase follows a minimum-dependency ethos in the interests of both longevity and decoupling from external APIs. It is for this reason \textsc{Octavius} defines its own snapshot parsing system which is not based on \texttt{yt}, as in \textsc{Caesar}. \textsc{Octavius} currently uses six dependencies which are all prominent, well-supported packages with active contributors: \textsc{h5py}, \textsc{numpy} \citep{harrisArrayProgrammingNumPy2020}, \textsc{astropy}, \textsc{numba}, \textsc{scipy} \citep{virtanenSciPy10Fundamental2020}, and \textsc{mpi4py}. \textsc{mpi4py} is listed as a dependency purely so package managers can handle \texttt{MPI} installations on clusters; \textsc{Octavius} fully supports serial analysis.

\section{Demonstration}

In this section we demonstrate the capabilities of the code on the fiducial \textsc{SIMBA} simulation run with 3D FOF haloes. This run has a box size of $100 \mathrm{ \, c \, Mpc \, h^{-1}}$ and $2 \times 1024^3$ resolution elements, with respective initial gas and DM masses of $1.82 \times 10^7 M_{\sun}$ and $9.6 \times 10^7 M_{\sun}$; its snapshots are approximately 240 GB. A full description and analysis of the \textsc{SIMBA} simulation suite is presented in \citet{dave_SIMBA}. We present a cross-code galaxy finder comparison to \textsc{Caesar} to validate this core feature of the code. Furthermore, we present standard astronomical scaling relations obtained from \textsc{Octavius}. 

The analysis was produced from a single machine with 24 \textit{Intel(R) Xeon(R) CPU E5-2697 v2 @ 2.70GHz} cores and 256 GB of RAM. Unless specified otherwise, this was done in an eight-rank configuration where each rank was allocated three threads and the default configuration file options were used.

\subsection{Validation} \label{sec:validation}

The results produced by \textsc{Octavius} strongly rely on its ability to accurately identify galaxies. To demonstrate this we use the galaxy stellar mass function (GSMF), a fundamental relation which is well-described in astrophysics literature across cosmic time. The GSMF serves as a means of validating the galaxy finder by the most basic property of galaxies: their stellar mass. For both toolkits, galaxies are defined as those above a resolution floor of 32 star particles. Observations are included for context; our goal here is not to have the model reproduce the observations, but rather to have \textsc{Octavius} reproduce the \textsc{Caesar} results.

\begin{figure*}
 \includegraphics[width=\textwidth]{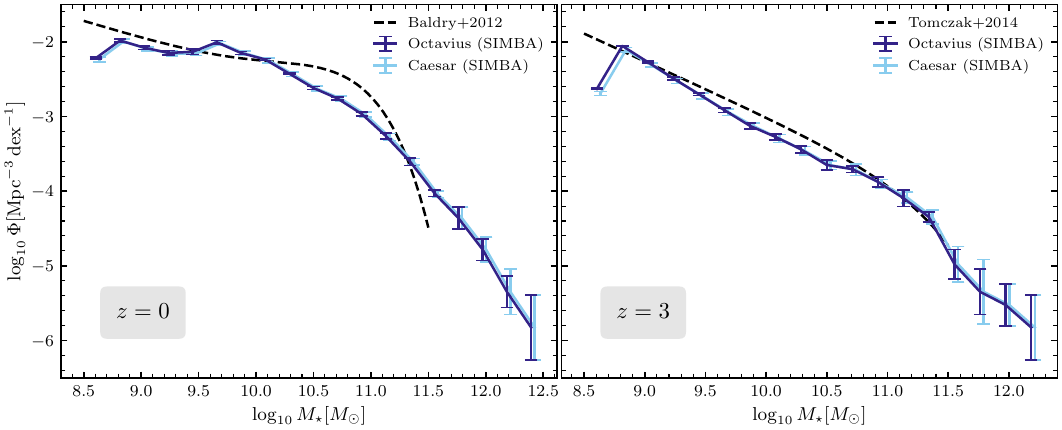}
 \caption{The GSMF at $z = 0$ and $z = 3$ produced from both \textsc{Caesar} and \textsc{Octavius} with contextual comparison to observations from \citet{baldryGalaxyMassAssembly2012} \& \citet{tomczakGalaxyStellarMass2014} respectively. The errors are computed from jackknife resampling of the eight simulation octants. To better illustrate the differences, the \textsc{Caesar} bins have been shifted 0.03 dex to the right to avoid an exact overlay.}
 \label{fig:gsmf}
\end{figure*}

\begin{table}
 \caption{The number of galaxies identified between \textsc{Caesar} and \textsc{Octavius} at $z = 0$. Note that both toolkits apply a minimum star criterion of 16 which is reflected here.}
 \label{tab:fof6d_difference}
 \begin{tabular}{lcc}
  \hline
  Toolkit & $N_{\rm galaxies}$ & $N_{\rm galaxies, resolved}$\\
  \hline
  \textsc{Octavius} & 67,526 & 48,162\\
  \textsc{Caesar} & 66,274 & 46,813\\
  \hline
 \end{tabular}
\end{table}

\Cref{fig:gsmf} shows the GSMF at $z = 0$ and $z = 3$. The results produced by \textsc{Octavius} mostly agree with its predecessor. However, though minor, there is an evident difference in the GSMF where \textsc{Octavius} generally identifies more galaxies above the resolution floor; this is more visible at $z = 3$. This is attributed to a discovered bug in the normalisation of the kernel in the velocity criterion of the \textsc{Caesar} FOF6D algorithm which caused structure to over-fragment. We include its effect here for comparison to prior published results with \textsc{Caesar} which its users are most familiar with. Over-fragmented galaxies are either pushed below the minimum star criterion, causing them to be excluded from the catalogue entirely, or fall below the resolution floor. This systematically biases the GSMF downwards. Referring to \cref{tab:fof6d_difference}, \textsc{Caesar} identifies fewer galaxies overall and loses more galaxies to the resolution cut. This is consistent with its galaxies being smaller due to overfragmentation. Examining \cref{fig:gsmf}, the effect is evidently more pronounced at higher redshifts where galaxies are smaller. This issue was first identified during development of \textsc{Octavius} and investigated before being fixed; it has since also been retroactively patched in \textsc{Caesar}.

\begin{figure*}
    \centering
    \includegraphics[width=\textwidth]{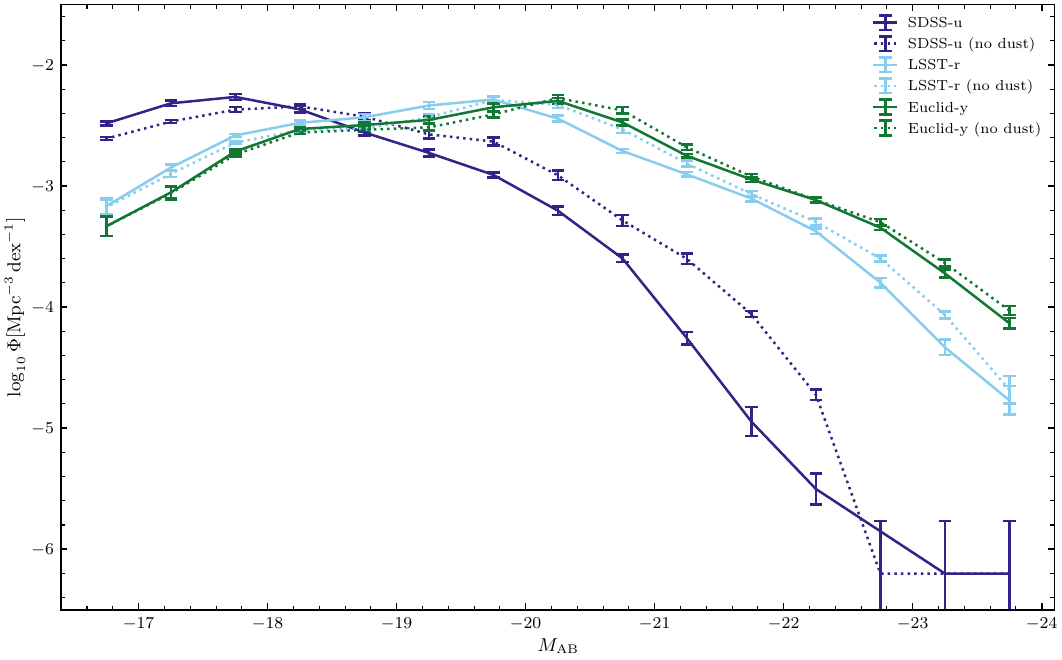}
    \caption{Luminosity functions in the \textit{SDSS-u}, \textit{LSST-r}, and \textit{Euclid-y} filters at $z = 0$, both with (solid line) and without (dotted line) dust. The dust-attenuated curves use the dust masses from the snapshot directly; we have assumed a \citet{chabrierGalacticStellarSubstellar2003} IMF. The turnover at the faint end is an artefact of the resolution floor; observed luminosity functions rise at the faint end. Error bars are computed from jackknife resampling of the eight simulation octants.}
    \label{fig:luminosity_funcs}
\end{figure*}

We additionally present results from the photometry routine. For this we examine three filters with and without dust attenuation: \textit{SDSS-u}, which is in the UV regime; \textit{LSST-r}, which is in the optical regime; and \textit{Euclid-y}, which is in the near-IR regime. \cref{fig:luminosity_funcs} shows the corresponding luminosity functions at $z = 0$. The \textsc{Caesar} routine contained minor bugs in its application of the Doppler shift, the kernel-weighted integral, and handling of periodic boundary conditions. In conjunction with the discussed bug in its galaxy finder, a direct comparison would be misleading as the results are not like-for-like. 

Attenuation can be seen to generally decrease the number of galaxies reaching higher-luminosity bins and shift the curve towards the faint end; the wavelength dependence of attenuation is visible in the more pronounced impact of dust on the bluer filters. The dust model therefore produces results which are physically consistent and sensible. 

The sharp decline in the \textit{SDSS-u} band at the bright end is consistent with the expectation the most massive galaxies are quenched: they are thus composed of older, redder stars which appear fainter in the near-UV. 

The benefit of this routine is in its speed: \textsc{Octavius} can produce reliable (though approximate) mock photometry for a snapshot significantly faster than heavyweight codes which handle full radiative transfer, including for simulation suites which do not track dust information.

\begin{figure*}
    \centering
    \includegraphics[width=\textwidth]{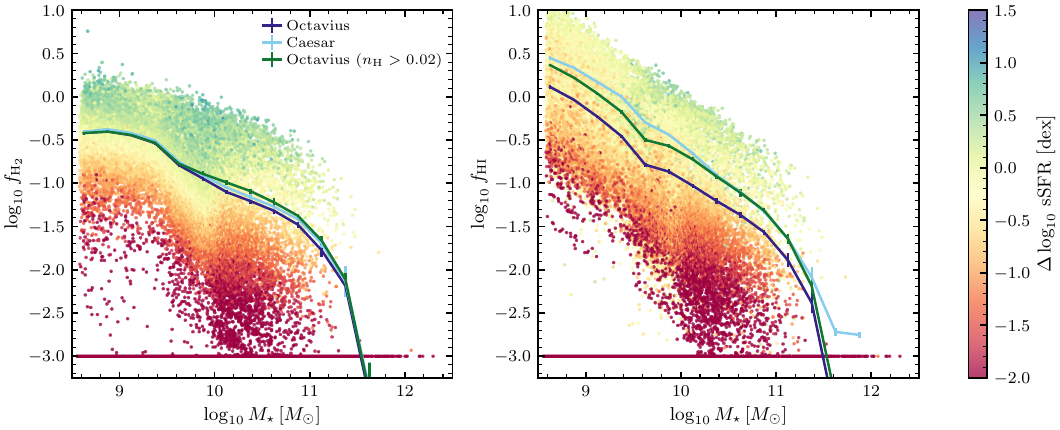}
    \caption{Molecular (left) and neutral (right) gas fractions compared between \textsc{Octavius} and \textsc{Caesar}. Points represent \textsc{Octavius} galaxies colour-coded by their deviation from the star-forming main sequence; lines represent running means. Errors are computed from jackknife resampling of the eight simulation octants. Galaxies with fraction $0$ are included at $-3$.}
    \label{fig:gas_fracs}
\end{figure*}

\Cref{fig:gas_fracs} shows the obtained hydrogen mass fractions from both toolkits. While both recover the same trend, there is a systematic offset between the toolkits under default settings. This is not primarily attributed to the aforementioned differences in the galaxy finding. In fact, the discrepancy arises from a decision in \textsc{Caesar} to automatically distribute the halo $\rm HI$ and $\rm H_{2}$ masses amongst their galaxies: the contribution from each gas particle is redistributed to its closest galaxy by mass-weighted distance. While this is a physically motivated decision to consider low-density gas, this step also has the effect of adding CGM and intracluster mass to galaxies. 

By contrast, the galaxy gas criterion is fully configurable in \textsc{Octavius}. To recover a similar result, users can simply tune parameters to allow low-density gas to be assigned to galaxies. We demonstrate this in~\cref{fig:gas_fracs} with an overlay of the running mean recovered by \textsc{Octavius} with a more permissive criterion. This approach also assigns the gas particles to the galaxies rather than adding their masses \textit{post hoc}, meaning they correctly propagate into dependent properties such as the dust attenuation. The enhanced flexibility of \textsc{Octavius} is clear.

\subsection{Performance} \label{sec:performance}

Whilst profiling the performance of the package, the machine was affected by strongly variable and degraded I/O bandwidth resulting from remedial work on the cluster. Stable I/O periods were brief; otherwise, the time spent in I/O could differ by over an order of magnitude between runs. We therefore focus our profiling on compute time scaling and memory usage. Nonetheless, the best observed end-to-end run of the pipeline on the fiducial \textsc{SIMBA} snapshot took $24.8$ minutes, with I/O accounting for $\sim$ 75 per cent of the total time. The pipeline is hence strongly I/O-bound; we acknowledge the inability to accurately profile the decomposition as a limitation of the results.

\begin{figure*}
    \includegraphics[width=\textwidth]{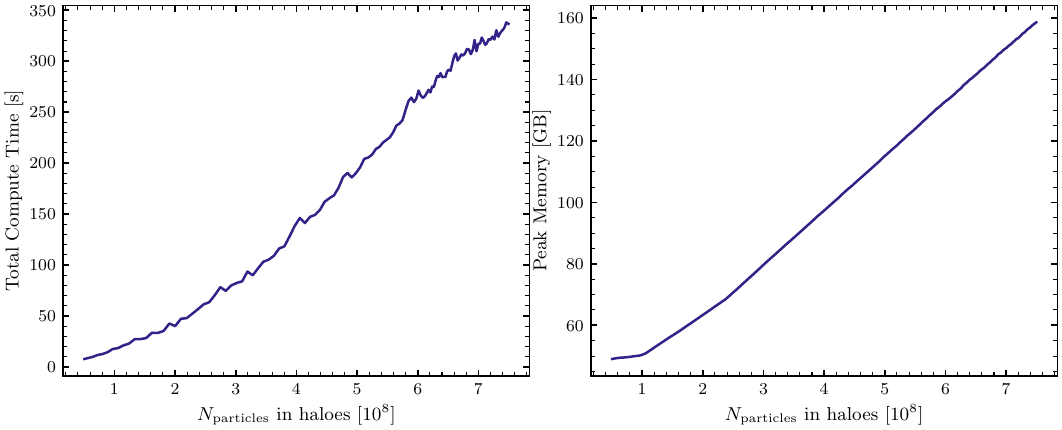}
    \caption{The total compute time and peak memory across snapshots as a function of the number of particles in haloes, which is a proxy for the total work. The curvature of runtime is consistent with per-halo work being superlinear (\cref{sec:parallelism}) and the formation of proportionally more massive haloes over cosmic time. Data were collected by sweeping from $z = 6.84$ to $z = 0$ across the fiducial \textsc{SIMBA} run. Peak memory is calculated by summing the individual peak memories of each rank, which do not necessarily coincide: thus, it is an upper bound.}
    \label{fig:scaling}
\end{figure*}

\Cref{fig:scaling} shows the scaling of total compute time and peak memory as the number of halo-assigned particles increases across the simulation. The total compute time is a mere $\sim$ 6 minutes on the largest snapshot. Compute can be seen to scale slightly superlinearly, consistent with the superlinear halo weight model. The snapshots comfortably fit within memory, a crucial result which suggests \textsc{Octavius} is well-prepared to tackle terabyte-scale snapshots. 

At the same time, the current data loading mechanism must load a slab of each snapshot dataset into memory temporarily (\cref{sec:parallelism}), regardless of how many particles are assigned to haloes. The I/O time should therefore remain constant with respect to the number of particles, depending only upon the file size: a caveat to the compute scaling. 

A direct comparison against \textsc{Caesar} on the same snapshot was not possible;  \textsc{Caesar} exceeded the RAM limit (256 GB) of the machine. To provide a like-for-like comparison on the same hardware, we ran an end-to-end analysis with all routines on a smaller \textsc{SIMBA} $50 \mathrm{ \, c \, Mpc \, h^{-1}}$ box at $z = 0$ during a period of stable I/O bandwidth. \textsc{Caesar} was allocated all 24 cores via \texttt{OpenMP}.

\begin{table}
 \caption{The runtime and peak memory of \textsc{Octavius} against \textsc{Caesar} on the smaller $50 \mathrm{ \, c \, Mpc \, h^{-1}}$ box at $z = 0$.}
 \label{tab:50box_comparison}
 \begin{tabular}{lcc}
  \hline
  Toolkit & Runtime [min] & Peak Memory [GB]\\
  \hline
  \textsc{Octavius} & 10.3 & 22.1 \\
  \textsc{Caesar} & 297.3 & 119.2\\
  \hline
 \end{tabular}
\end{table}

\Cref{tab:50box_comparison} details the results of this comparison: \textsc{Octavius} shows an order-of-magnitude improvement in speed over \textsc{Caesar} with a factor of five reduction in peak memory. Furthermore, as \textsc{Octavius} uses \texttt{MPI} to parallelise its analysis of an individual snapshot, it is not bounded to the RAM of a single node like \textsc{Caesar} is. 

\begin{figure}
    \centering
    \includegraphics[width=\columnwidth]{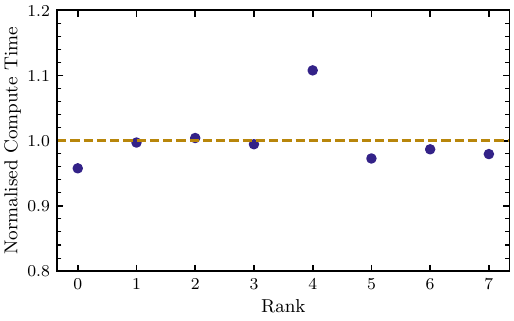}
    \caption{The runtime balance across eight \texttt{MPI} ranks on the $z = 0$ snapshot of the fiducial \textsc{SIMBA} simulation, normalised to the mean.}
    \label{fig:rank_balance}
\end{figure}

Returning to the fiducial run, \cref{fig:rank_balance} shows the runtime balance across ranks. Evidently, the simple weight laws (\cref{sec:parallelism}) produce a well-balanced computational load: the parallel efficiency is 89 per cent. The outlying rank 4 is likely a result of a galaxy-level imbalance which the halo-level distribution of work cannot account for.

\begin{figure}
    \centering
    \includegraphics[width=\columnwidth]{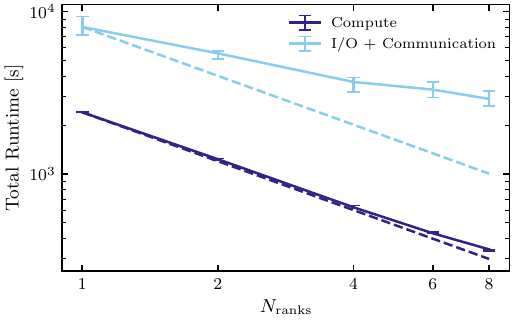}
    \caption{The strong scaling, measured by increasing the number of \texttt{MPI} ranks (three threads/rank). We include I/O in this plot: in absolute terms the bandwidth was degraded but stable, so the relative scaling is representative. Dashed lines represent theoretical perfect scaling: $T = T_0 \, / \, R$, where $T_0$ is the runtime of one rank and $R$ is the number of ranks.}
    \label{fig:strong_scaling}
\end{figure}

Finally,~\cref{fig:strong_scaling} shows the strong scaling as a function of \texttt{MPI} ranks. Compute approaches near-perfect scaling; I/O and communication only see a minor gain, which is expected as the cluster uses an NFS filesystem. The parallel reads described in~\cref{sec:parallelism} are better-suited to a Lustre filesystem, which is purpose-built for parallel I/O and HPC environments.  It is worth noting that in the current setup, \textsc{Octavius} is I/O-limited.

\section{Discussion, Summary, and Future Work}

We have introduced \textsc{Octavius}, a high-performance analysis toolkit for post-processing galaxy simulations. \textsc{Octavius} possesses the level of scalability required to efficiently analyse the next generation of simulations, which are expected to exacerbate the big data computational challenge faced by analysis software. It achieves this through fast, optimised \textsc{numba} code; an HPC-friendly data model; efficient algorithms; and simple yet effective halo-based parallelism. The output catalogues are agnostic to the input simulation format, enabling cross-code comparison through snapshot parsers. We have demonstrated the performance of the code and presented physics results contrasted with its predecessor, \textsc{Caesar}, to whom \textsc{Octavius} remains faithful whilst delivering order-of-magnitude performance improvements. Crucially, the memory footprint is significantly reduced: the memory requirements of \textsc{Caesar} have grown untenable for large snapshots. \textsc{Octavius} will therefore enable analyses which were not possible with its predecessor.

We have also described the flexible, extensible architecture of the code: this includes important, novel features over \textsc{Caesar} such as unit testing, support for processing subhaloes, and a configuration file. The modularised architecture and strict separation of responsibilities throughout the codebase have simplified future additions. \textsc{Octavius} is designed with an \textit{ab initio} expectation it will grow over the coming years, in light of the unplanned evolution of \textsc{Caesar}. The comprehensive unit and regression testing framework will further support the long-term evolution of the codebase.

The package is intended to be accessible to a diverse user base, from those running simple \textsc{Python} scripts on small snapshots to large-scale analysis sweeps on supercomputing clusters. The API facilitates both use cases and provides a convenient means of interfacing with the output HDF5 catalogues. The catalogues are comprehensive and enable users to explore the relationships between haloes, subhaloes, and galaxies; the particle index mappings in the catalogue can be used to compute bespoke properties from the original snapshot. The \texttt{OctaviusAnalyser} enables more detailed and tailored analyses beyond catalogue-level properties.

In its current state, \textsc{Octavius} supports five simulation suites. However, the analysis has mostly been stress-tested on \textsc{SIMBA} snapshots: other suites have been primarily tested using snapshots from smaller boxes. Following the public release we hope the code will be similarly stress-tested at scale for a variety of simulations. This will provide useful diagnostics such as whether the empirical load-balancing power laws hold across suites at scale. We expect this to be the case because of the underlying physics, such as the halo mass function; however, the different membership assignments made by different halo finders may challenge this assumption. 

The main limitation of the code at present is its I/O cost. The redistribution of particles to ranks according to halo membership requires each rank to read a contiguous slice of the full dataset followed by an \texttt{MPI\_Alltoallv} call. While the analysis is still very fast, we are nonetheless considering ways of reducing the I/O footprint of the pipeline. One potential approach would be to reorder the snapshot on disc such that each halo's particles are contiguous. There is precedent for this in \textsc{TNG} snapshots with associated \textsc{SUBFIND} catalogues. Not only would this reduce I/O costs by avoiding reading the full dataset, it would also enable the pipeline to implement chunking for out-of-core computing. However, this would require modifying the snapshot itself which would have consequences for other analysis scripts expecting an unaltered format. Solutions to the I/O problem are generally not straightforward, as the layout of data on disc differs between simulation suites and \textsc{Octavius} is a multi-suite toolkit; all formats follow the same I/O path from the \texttt{SnapshotReader} base. To improve write speeds for the catalogue, a parallel HDF5 build is under consideration.

There is an existing ecosystem for analysing simulation data which \textsc{Octavius} will complement. \textsc{SOAP}, as described in \citet{mcgibbonSOAPPythonPackage2025}, is a spatially based catalogue builder for \textsc{SWIFT} snapshots: it computes properties within apertures placed on the centres of externally identified haloes. \textsc{SOAP} has a similar science goal to \textsc{Octavius} of computing properties, but only supports \textsc{SWIFT} and does not include galaxy finding. Its particle-level spatial approach means its apertures do not face the reduced accuracy of the halo-local apertures in \textsc{Octavius}. \textsc{yt} \citep{turkYtMulticodeAnalysis2011} and \textsc{pynbody} \citep{pontzenPynbodyNBodySPH2013} are multi-code frameworks for interfacing with particle-level snapshot data. They are less pipeline-oriented, operating more akin to the \texttt{OctaviusAnalyser}, and place a heavier emphasis on visualisation; the previous \textsc{Caesar} pipeline was an extension to \textsc{yt}. \textsc{SWIFTGalaxy} \citep{omanSWIFTGalaxyPythonPackage2025} is a package for analysing externally identified galaxies in snapshots at the particle level, and is currently being extended to work with galaxies identified by \textsc{Octavius}.

\textsc{Octavius} is expected to replace \textsc{Caesar} as an important component of the software stack of multiple research groups. The codebase was designed with longevity in mind, with the declarative architecture and \textsc{Python}-first ethos intended to support future development. To this end, \textsc{Octavius} remains in active development, with new features such as datacubes and line absorption spectra currently in progress.

\section*{Acknowledgements}

We thank Kyle Oman (\textit{Durham University}) for providing \textsc{SWIFT} snapshots, posing ideas, and providing useful feedback on the code. We thank Weiguang Cui (\textit{Universidad Aut\'{o}noma de Madrid}) for providing reference \textsc{Caesar} and \textsc{AHF} catalogues. We thank Hugo Robijns (\textit{University of Cambridge}) for useful and thought-provoking discussion regarding the code logic. We thank Kevin Lo (\textit{University of Edinburgh}) for providing advice on halo finders. Finally, we are grateful to the many open-source software developers upon whose work \textsc{Octavius} is based. This work was supported in part by the Polish National Science Centre grant 2024/53/B/ST9/03747.

During development, Claude Opus 4.6 (Anthropic) was used in an advisory capacity for code review and advice. All source code, unit tests, and documentation were written directly by the authors, who take full responsibility for both the software and the content of this manuscript.

\section*{Data Availability}

The \textsc{Octavius} source code and its history are publicly and freely available on GitHub at \url{https://github.com/jp-duminy/octavius}; production releases are archived on Zenodo with the DOI 10.5281/zenodo.22166418 \citep{duminyOctavius2026}.

\section*{Conflict of Interest}

Authors declare no conflict of interest.


\bibliographystyle{rasti}
\bibliography{bibliography} 



\appendix
\section{Supported Codes}

\Cref{tab:suites} lists the simulation suites and respective halo finders supported by \textsc{Octavius}.

\begin{table*}
 \caption{The simulation suite and halo finder combinations supported by \textsc{Octavius}.}
 \label{tab:suites}
 \begin{tabular*}{\textwidth}{l@{\hspace*{50pt}}l@{\hspace*{50pt}}}
  \hline
  Suite & Halo Finders\\
  \hline
  \noalign{\vskip 4pt}
  \textsc{SIMBA} & SNAPSHOT, AHF \\
  \textsc{SWIFT-KIARA} & SNAPSHOT, AHF, HBT-HERONS\\
  \textsc{SWIFT-EAGLE} & SNAPSHOT, AHF, HBT-HERONS\\
  \textsc{SWIFT-COLIBRE} & SNAPSHOT, AHF, HBT-HERONS\\
  \textsc{TNG} & SNAPSHOT, AHF, SUBFIND\\
  \hline
 \end{tabular*}
\end{table*}

\section{Configuration File Parameters}

\Cref{tab:config} lists the relevant physics parameters in the YAML configuration file.

\begin{table*}
 \caption{The physics parameters of the configuration file.}
 \label{tab:config}
 \begin{tabularx}{\textwidth}{l >{\raggedright\arraybackslash}X r}
  \hline 
  Parameter & Description & Default\\
  \hline
  \noalign{\vskip 4pt}
  \texttt{min\_stars\_per\_galaxy} & Minimum number of star particles a collection of baryons needs to be considered a galaxy (6D FOF) & 16\\
  \texttt{min\_dm\_per\_halo} & Minimum number of DM particles a halo must have to be analysed & 24\\
  \texttt{nH\_lim} & Density ($\mathrm{n_H \, cm^{-3}}$) above which gas is considered dense & 0.13\\
  \texttt{T\_lim} & Temperature ($K$) above which gas is considered hot & 100,000\\
  \texttt{FRAD} & Radiative efficiency of accretion $\epsilon$ & 0.1\\
  \texttt{MU} & Mean molecular weight for gas (for virial relations) & 0.6\\
  \texttt{b} & 3D FOF $b$ parameter & 0.02\\
  \texttt{velocity\_factor} & The 6D FOF $\alpha$ parameter & 1.0\\
  \texttt{gas\_criterion} & Gas filter criteria for FOF & \texttt{COLD\_OR\_STARFORMING}\\
  \texttt{radial\_quantiles} & The quantile of mass for each enclosed mass profile (e.g. $r_{20}$) & \text{{$r_{20}$: 0.2, \texttt{half\_mass}: 0.5, $r_{80}$: 0.8}}\\
  \texttt{aperture\_size} & Size of aperture mass ($\rm c kpc$) & 30\\
  \texttt{virial\_factors} & Virial overdensity factors & \text{[200, 500, 2500]}\\
  \texttt{halo\_centre} & Reference position for halo centres & \texttt{MIN\_POT}\\
  \texttt{bands} & \textsc{FSPS}-supported bands to compute photometry in & \texttt{all}\\
  \texttt{extinction\_law} & Extinction law to apply to spectra & \texttt{COMPOSITE}\\
  \texttt{viewing\_axis} & LOS axis for photometry & \texttt{Z}\\
  \texttt{use\_dust} & Whether to use dust masses from the snapshot & \texttt{True}\\
  \texttt{use\_cosmic\_extinction} & Applies \citet{madauRadiativeTransferClumpy1995} IGM extinction & \texttt{True}\\
  \texttt{interpolation\_bins} & Number of bins to use in numerical integration of kernel LOS integral & 5000\\
  \texttt{kernel\_type} & Type of kernel used \citep{monaghan_sph} & \texttt{CUBIC} \\
  \texttt{power\_law\_alpha} & The exponent in power-law attenuation & 1.0\\
  \texttt{split\_age} & Age ($\rm Gyr$) below which a star particle's age is split into sub bins & 0.01\\
  
  \hline
 \end{tabularx}
\end{table*}

\section{Output Properties}

\Cref{tab:properties} lists the properties output by each stage.

\begin{table*}
\caption{The output properties of all currently implemented analysis routines.}
\label{tab:properties}
    \begin{tabularx}{\textwidth}{l c >{\raggedright\arraybackslash}X}
    \hline
    Property & Symbol & Description \\
    \hline
    \noalign{\vskip 4pt}
    \multicolumn{3}{l}{\textit{Core Properties}} \\[3pt] 
    \texttt{com\_pos} & $\vec{x}_{\rm COM}$ & Centre of mass position vector\\
    \texttt{com\_vel} & $\vec{v}_{\rm COM}$ & Centre of mass velocity vector\\
    \texttt{L} & $\vec{L}$ & Angular momentum vector\\
    \texttt{inertia\_tensor} & $\mathbf{I}$ & Inertia tensor\\
    \texttt{velocity\_dispersion} & $\sigma$ & Mass-weighted 3D velocity dispersion\\
    \texttt{radius\_\{quantile\}} & $r_{\rm quantile}$ & Enclosed mass radii \\
    \texttt{radius\_max} & $r_{\rm member, max}$ & Maximum member radius\\
    \texttt{mass} & $M$ & Total mass\\
    \texttt{n} & $N$ & Total number of particles\\[4pt]
    \multicolumn{3}{l}{\textit{Core Properties (Haloes)}} \\[4pt]
    \texttt{r200m} & $r_{200m}$ & Radius enclosing 200x the mean matter density\\
    \texttt{spin\_param} & $\lambda$ & Bullock spin parameter \citep{bullockUniversalAngularMomentum2001}\\
    \texttt{temperature\_virial} & $T_{\rm vir}$ & Virial temperature\\
    \texttt{velocity\_circular} & $v_{\rm circ}$ & Circular velocity at $r_{200m}$\\
    \texttt{vmax} & $v_{\rm max}$ & The maximum circular velocity\\
    \texttt{rmax} & $r_{\rm max}$ & The radius at $v_{\rm max}$.\\
    \texttt{r\{factor\}c} & $r_{\rm factor \times c}$ & Radius enclosing the critical density factor e.g. $r_{200c}$\\
    \texttt{m\{factor\}c} & $M_{\rm factor \times c}$ & Mass enclosed within $r_{\rm factor \times c}$\\
    \texttt{minpot\_pos} & $\vec{x}_{\Phi_{\rm min}}$ & Position vector of the particle at the minimum potential\\
    \texttt{minpot\_vel} & $\vec{v}_{\Phi_{\rm min}}$ & Velocity vector of the particle at the minimum potential\\[4pt]
    \multicolumn{3}{l}{\textit{Core Properties (Galaxies)}} \\[4pt]
    \texttt{BoverT} & $B / T$ & Bulge-to-total kinematic ratio\\
    \texttt{kappa\_rot} & $\kappa_{\rm rot}$ & Ratio of rotational-to-total kinetic energy\\[4pt]
    \multicolumn{3}{l}{\textit{Particle-Type-Specific Properties}} \\[4pt]
    \texttt{mass\_HI} & $M_{\rm HI}$ & Neutral hydrogen mass\\
    \texttt{mass\_H2} & $\rm M_{H_{2}}$ & Molecular hydrogen mass\\
    \texttt{sfr} & $\rm SFR$ & Star formation rate\\
    \texttt{metallicity\_gas\_mass\_weighted} & $\langle Z_{\rm M,gas} \rangle $ & Mass-weighted gas metallicity\\
    \texttt{metallicity\_gas\_sfr\_weighted} & $\langle Z_{\rm SFR} \rangle $ & SFR-weighted gas metallicity\\
    \texttt{temperature\_gas\_mass\_weighted} & $\langle T_M \rangle $ & Mass-weighted gas temperature\\
    \texttt{mass\_gas\_hot} & $M_{\rm hot}$ & Hot gas mass\\
    \texttt{mass\_gas\_cold} & $M_{\rm cold}$ & Cold gas mass\\
    \texttt{metallicity\_star\_mass\_weighted} & $\langle Z_{M, \star} \rangle $ & Mass-weighted stellar metallicity\\
    \texttt{age\_star\_mass\_weighted} & $\langle \tau_M \rangle $ & Mass-weighted mean stellar age\\
    \texttt{age\_star\_metal\_weighted} & $\langle \tau_Z \rangle $ & Metallicity-weighted mean stellar age\\
    \texttt{smbh\_mass} & $M_{\rm SMBH}$ & Mass of most massive black hole\\
    \texttt{smbh\_fedd} & $\lambda_{\rm SMBH}$ & Eddington fraction of most massive black hole\\
    \texttt{smbh\_mdot} & $\dot{M}_{\rm SMBH}$ & Accretion rate of most massive black hole\\[4pt]
    \multicolumn{3}{l}{\textit{Particle-Type-Specific Properties (Haloes)}} \\[4pt]
    \texttt{mass\_cgm} & $M_{\rm CGM}$ & CGM gas mass\\
    \texttt{metallicity\_gas\_mass\_weighted\_cgm} & $\langle Z_{\rm M,CGM} \rangle $ & Mass-weighted CGM gas metallicity\\
    \texttt{metallicity\_gas\_temperature\_weighted\_cgm} & $\langle Z_{\rm T, CGM} \rangle $ & Temperature-weighted CGM gas metallicity\\
    \texttt{temperature\_gas\_mass\_weighted\_cgm} & $\langle T_{\rm M, CGM} \rangle $ & Mass-weighted CGM gas temperature\\
    \texttt{temperature\_gas\_metal\_weighted\_cgm} & $\langle T_{\rm Z, CGM} \rangle $ & Metallicity-weighted CGM gas temperature\\[4pt]
    \multicolumn{3}{l}{\textit{Local Environment Properties}} \\[4pt]
    \texttt{mass\_\{aperture\}kpc} & $M_{\rm aperture}$ & Mass in specified aperture size\\
    \texttt{mass\_HI\_\{aperture\}kpc} & $M_{\rm HI, aperture}$ & $\rm HI$ mass in specified aperture size\\
    \texttt{mass\_H2\_\{aperture\}kpc} & $M_{\rm {H_2}, aperture}$ & H$_{2}$ Mass in specified aperture size\\[4pt]
    \multicolumn{3}{l}{\textit{Photometry}} \\[4pt]
    \texttt{mag\_abs\_\{band\}} & $M$ & Absolute magnitude in specified band\\
    \texttt{mag\_app\_\{band\}} & $m$ & Apparent magnitude in specified band\\  
    \texttt{mag\_abs\_nodust\_\{band\}} & $M_{\rm no \, dust}$ & Absolute magnitude in specified band (no dust attenuation) \\
    \texttt{mag\_app\_nodust\_\{band\}} & $m_{\rm no \, dust}$ &Apparent magnitude in specified band (no dust attenuation) \\  
    \texttt{luminosity\_fir} & $L_{\rm FIR}$ & Dust-reprocessed luminosity\\
    \texttt{beta} & $\beta$ & UV spectral slope\\
    \texttt{beta\_nodust} & $\beta_{\rm no \, dust}$ & UV spectral slope (no dust attenuation)\\
    \texttt{spectra} & $F_\lambda$ & Observed spectra\\
    \texttt{spectra\_nodust} & $F_{\lambda, \rm no \, dust}$ & Observed spectra (no dust attenuation)\\

    \hline
    \end{tabularx}
\end{table*}

\section{Documentation and Outputs}
A complete list of the current output properties is available in the documentation, which is available at https://octavius.readthedocs.io/en/latest/index.html. The documentation for the present version is archived on Zenodo (DOI 10.5281/zenodo.22166418,  \citealt{duminyOctavius2026}).


\bsp	
\label{lastpage}
\end{document}